\documentstyle[12pt]{article}
\newcommand{\bee}{\begin{equation}}
\newcommand{\ee}{\end{equation}}
\newcommand{\ba}{\begin{array}}
\newcommand{\ea}{\end{array}}
\newcommand{\bea}{\begin{eqnarray}}
\newcommand{\eea}{\end{eqnarray}}

\newcommand{\R}{\rm I\kern-.2emR}
\newcommand{\Z}{{\rm Z\kern-.35em Z}}
\begin{document}
\thispagestyle{empty}
\begin{flushright}
MPI-PhT/95-50\\
June 1995
\end{flushright}
\bigskip\bigskip\bigskip\begin{center}
{\LARGE {Lattice fermions with gauge}}
\vskip 7pt
{\LARGE {noninvariant measure}}
\end{center}
\vskip 1.0truecm
\centerline{
Sergei V. Zenkin\footnote{Permanent address: Institute for Nuclear
Research of
the Russian Academy of Sciences, 117312 Moscow, Russia\\
E-mail address: zenkin@inr.msk.su}}
\vskip5mm
\centerline{Max-Planck-Institut f\"ur
 Physik}
\centerline{ -- Werner-Heisenberg-Institut -- }
\centerline{F\"ohringer Ring 6, 80805 Munich, Germany}
\vskip 2cm
\bigskip \nopagebreak \begin{abstract}
\noindent
We define Weyl fermions on a finite lattice in such a way that in the path
integral the action is gauge invariant but the functional measure is not.
Two variants of
such a formulation are tested in perturbative calculation of the fermion
determinant in chiral Schwinger model. We find that one of these variants
ensures restoring the gauge invariance of the nonanomalous part of the
determinant in the continuum limit. A `perfect' perturbative regularization
of the chiral fermions is briefly discussed.
\end{abstract}
\vskip 1.5cm

\newpage\setcounter{page}1


\section{Introduction}

The fermion determinant for the Weyl fermions is known to break gauge
invariance
producing the chiral anomalies (see, for example, \cite{Ano} and the
references therein). In the path integral formulation such breaking may
have two
origins: gauge noninvariance of the fermion action or the noninvariance
of the functional fermion measure \cite{Fu}. In all known formulations of
the fermions
on a lattice the fermion measure is defined to be gauge invariant and the
responsibility for the anomalies, together with all the well-known problems
of defining the chiral lattice fermions \cite{KS,NN}, is transfered to the
action (for the review  of the recent approaches and the references see, for
example \cite{Ref}).

In this paper we consider a formulation of the Weyl fermions on a finite
lattice in which the action is gauge invariant, and the anomaly originates
from the gauge noninvariance of the measure\footnote{
To our best knowledge the first mention of similar example of a lattice
gauge theory in the literature is the footnote 2 in the ref. \cite{BD}.
Recently, gauge-variant measure was discussed within the approach
\cite{NaN}, which however employs infinitely many fermionic degrees of
freedom.}.
Both the action and the measure are invariant under the global chiral
transformations.
The formulation employs additional
Grassmann variables,
which however are not dynamical since they are eliminated by a
constraint involved in the measure, and the gauge variables living on the
halfs of the lattice links.
The action now is determined uniquely, while the constraint
includes a certain ambiguity. Although such a formulation can still be
transformed by changing variables to one with the conventional measure,
it does not repeat the formulations already known, rather it gives one a
new outlook on them.

Here we limit ourselves to two-dimensional theories and consider in
detail two variants of the constraint. We test them in a perturbative
calculation of the fermion determinant in the chiral Schwinger model,
and demonstrate that
for smooth gauge fields both variants leads to the correct results in the
continuum limit. The remarkable fact is that one of them  in this limit ensures
a restoration of the gauge invariance of the nonanomalous part of the
determinant.

In sect. 2 we introduce the formulation and discuss the variants of the
constraint; in sect. 3  the changes of variables in the path integral leading
it to more conventional forms are considered;
the calculation of the fermion
determinant is outlined in sect. 4; sect. 5 contains the summary and
discussion of a `perfect' perturbative regularization of the Weyl fermions.

Our conventions are the following: we consider square regular
lattice $\Lambda$ with spacing $a = 1$ and the size $N \times N$,
where $N$
is even; its sites numbered by $n = (n_0, n_1)$, $-N/2+1 \leq n_{\mu}
\leq N/2$; $\hat{\mu} = (\hat{0}, \hat{1})$ are the unit vectors along the
lattice links in the positive directions, so that, for instance, notation
$n + \frac{1}{2} \hat{0}$ means the middle of the link $(n, n+\hat{0})$.
We shall define the theory on a torus $S^1 \times S^1$ which is obtained by
the addition of links connecting each site $(N/2, \: n_1)$ with the site
$(-N/2+1, \: n_1)$ and the site $(n_0, \: N/2)$ with $(n_0,\: -N/2+1)$.

\section{Formulation}

Consider first free two-dimensional Weyl fermions whose action
in the continuum Euclidean space-time reads as
\bea
S &=& \int d^2 x \; \chi^*(x)\: (\partial_0 + i \partial_1) \:\chi(x) \cr
  &=& \int \frac{d^2 p}{(2 \pi)^2} \; \chi^*(p)\: i\: (p_0 + i p_1)\:
\chi(p)
\eea
(`left-handed' fermions). Despite the extreme simplicity, this action cannot
be transcribed on a lattice unless certain reasonable conditions
(which we want to be fulfilled) are violated \cite{NN}.

One of the origins of the problem is that there are no elements of
a lattice adequate to the non-tensorial nature of the
variables $\chi$
and $\chi^*$
\footnote{These variables are transformed under rotation by angle
$\phi$ as $\chi'(x') = \exp (-i \phi/2) \chi(x)$, ${\chi^*}'(x') =
\exp (-i \phi/2) \chi^*(x)$, i.e. as a `square root of a complex vector'.},
on which these variables might be defined in accordance with
arguments of homology theory \cite{Ra}.
Therefore, let us introduce the Grassmann variables $\chi^{*}_{n}$ and
$\chi^{0}_{n + \frac{1}{2} \hat{0}}$, $\chi^{1}_{n + \frac{1}{2} \hat{1}}$
defined on the lattice sites and the links, respectively, and define the
following local form invariant under the lattice translations and rotations:
\bea
A[\chi^{*}, \chi^{0}, \chi^{1}] &=& \sum_{n \in \Lambda}
\chi^{*}_{n} \left[\chi^{0}_{n + \frac{1}{2}
\hat{0}}
- \chi^{0}_{n - \frac{1}{2} \hat{0}} + i\: (\chi^{1}_{n +
\frac{1}{2} \hat{1}} - \chi^{1}_{n - \frac{1}{2} \hat{1}} )\right] \cr
  &=& \frac{1}{N^2} \sum_{p \in \Lambda^*} \chi^{*}_{p} \:i\:
\left[ 2 \sin (\frac{1}{2}p_0)
\: \chi^{0}_{p} + i \: 2 \sin (\frac{1}{2}p_1) \: \chi^{1}_{p} \right].
\eea
The variables $\chi^{*}_{n}$,
$\chi^{0}_{n + \frac{1}{2} \hat{0}}$, and $\chi^{1}_{n + \frac{1}{2}
\hat{1}}$ obey antiperiodic boundary conditions, but the variables
$\chi^{*}_{p} = \sum_{n \in \Lambda} \exp (i p n) \chi^{*}_{n}$,
$\chi^{0}_{p} = \sum_{n \in \Lambda} \exp [-i p (n + \frac{1}{2} \hat{0})]
\chi^{0}_{n + \frac{1}{2} \hat{0}}$, and $\chi^{1}_{p} = \sum_{n \in \Lambda}
\exp [-i p (n + \frac{1}{2} \hat{1})]
\chi^{1}_{n + \frac{1}{2} \hat{1}}$
defined
on the momentum lattice $\Lambda^*$, which topologically also is a torus,
\footnote{By virtue of the antiperiodic
boundary conditions on $\Lambda$, the momenta $p_{\mu}$ take the values
$p_{\mu} =
2 \pi (k_{\mu} - \frac{1}{2})/N$, where $-N/2+1 \leq k_{\mu}
\leq N/2$.} obey different boundary conditions:
\bea
\chi^{*}_{p + 2 \pi \hat{\mu}} &=& \chi^{*}_{p}, \cr
\chi^{0}_{p + 2 \pi \hat{0}} &=& -\chi^{0}_{p}, \quad
\chi^{0}_{p + 2 \pi \hat{1}} = \chi^{0}_{p}, \cr
\chi^{1}_{p + 2 \pi \hat{1}} &=& -\chi^{1}_{p}, \quad
\chi^{1}_{p + 2 \pi \hat{1}} = \chi^{0}_{p}.
\eea

We are aiming to define the Weyl fermions on $\Lambda$ by $2 \times N^2$
dynamical variables. In order to eliminate the superfluous $N^2$ variables
in $A$, impose on the $\chi^{0}$ and $\chi^{1}$ a linear constraint
\bee
F^{0}_{n} [\chi^0] - F^{1}_{n} [\chi^1] = 0, \quad \det F^{\mu} \neq 0.
\ee
It is naturally to limit the consideration to such $F$ that in the momentum
space take the diagonal form:
\bee
f^{0}(p) \: \chi^{0}_{p}  - f^{1}(p) \: \chi^{1}_{p} = 0, \quad
f^{\mu}(p) \neq 0, \;  p_{\mu} \in (-\pi,\: \pi),
\ee
with $f^{\mu}(p)$ being real. Of course the constraint must not break the
symmetry
of the lattice, in
particular the functions $f^{\mu}$  must be compatible with the
boundary conditions (3).

Obviously, the necessary conditions for the system (2), (5)
to define the `left-handed' Weyl fermions on the lattice are: positivity
of the product $f^0(p) f^1(p)$ within the Brillouin zone
$p_{\mu} \in (-\pi,\: \pi) $;
and $\lim_{N \rightarrow \infty, \: p N = const.} f^{\mu}(p) = 1$.
We shall not analyse the sufficient conditions for that. We only note that
the ambiguity in $f^{\mu}$ is closely related to the ambiguity in the
action in the conventional formulation and in the end of this  section
consider some explicit examples.

The action $A$ and the constraint (4) are invariant under the global
transformation $\chi^* \rightarrow \chi^* \: h^*$, $\chi^0
\rightarrow h \: \chi^0$, $\chi^1 \rightarrow h \: \chi^1$, where $h \in U(1)$
(the generalization to other groups is evident). Let us require the action
to be invariant under the gauge transformations
\bee
\chi^{*}_{n} \rightarrow \chi^{*}_{n} \: h^{*}_{n}, \quad
\chi^{0}_{n + \frac{1}{2} \hat{0}} \rightarrow h_{n + \frac{1}{2} \hat{0}}
\: \chi^{0}_{n + \frac{1}{2} \hat{0}}, \quad
\chi^{1}_{n + \frac{1}{2} \hat{1}} \rightarrow h_{n + \frac{1}{2} \hat{1}}
\: \chi^{1}_{n + \frac{1}{2} \hat{1}}.
\ee
This can be done by introducing the gauge variables defined on the
halfs of the links $U_{n, n + \frac{1}{2}\hat{\mu}}$,
$ U_{n, n - \frac{1}{2}\hat{\mu}} = U^{*}_{n - \frac{1}{2}\hat{\mu}, n}$,
with obvious gauge transformations\footnote{In fact these variables are a
variant of those used in ref. \cite{BZ}.}.
Then the gauged action (2) takes the form:
\bea
A[\chi^{*}, \chi^{0}, \chi^{1}; U]
= \sum_{n \in \Lambda} &\chi^{*}_{n}&
\left[ U_{n, n + \frac{1}{2}\hat{0}} \: \chi^{0}_{n + \frac{1}{2} \hat{0}}
-  U_{n, n - \frac{1}{2}\hat{0}} \: \chi^{0}_{n - \frac{1}{2} \hat{0}} \right.
\cr
&&\mbox{} + i\: \left. (  U_{n, n + \frac{1}{2}\hat{1}} \: \chi^{1}_{n +
\frac{1}{2} \hat{1}} -  U_{n, n - \frac{1}{2}\hat{1}} \:
\chi^{1}_{n - \frac{1}{2} \hat{1}} ) \right].
\eea

We now define the generating functional for the system (7), (4) as:
\bee
Z_{F}[U; \eta^*, \eta] =
\frac{1}{{\cal N}_{F}} \int d \mu_F
\exp \{ - A[\chi^{*}, \chi^{0}, \chi^{1}; U] + \sum_{n \in \Lambda} (
\eta^{*}_{n} \: F^{0}_{n}[\chi^{0}] + \chi^{*}_{n} \: \eta_{n} ) \},
\ee
where the measure has the form
\bee
d \mu_F  =
\prod_{n \in \Lambda} d \chi^{0}_{n + \frac{1}{2}
\hat{0}} \: d \chi^{1}_{n + \frac{1}{2} \hat{1}} \: d \chi^{*}_{n} \:
\delta(F^{0}_{n} [\chi^0] - F^{1}_{n} [\chi^1]),
\ee
${\cal N}_F$ is a normalization factor such that $Z_{F}[0; 0, 0] = 1$,
and $\eta^*$ and $\eta$ are external sources.

Thus, the action in this path integral is gauge invariant, while the measure
is invariant only under the global transformations.

{}From the form of the fermion propagator that immediately follows from (8),
\bee
S(p) = - i\: \left[ 2 \sin (\frac{1}{2} p_0) \: \frac{1}{f^{0}(p)}
 + i \: 2 \sin (\frac{1}{2}p_1) \frac{1}{f^{1}(p)} \right]^{-1},
\ee
it is seen that this approach does not guarantee to avoid the pathologies like
species doubling. For example the choice $f^{\mu}(p) = 1/\cos(\frac{1}{2}
p_{\mu})$ leads to the naive propagator. Although the coupling of the gauge
fields to the fermions in the action (7) leaves some of the undesirable
fermion modes decoupled, such cases should be avoided. Note that such a
choice of the constraint does not look
natural. Below we consider two examples of the constraint which in a certain
sense are the most natural.

The problem would be solved perfectly if one could put
$f^{\mu}(p) = 1$, but that is clearly impossible
by virtue of boundary conditions (3). However, one can still satisfy this
condition for the momenta within the Brillouin zone. In this case
\bee
f^{\mu}(p) = \epsilon (p_{\mu}),
\ee
where
the $\epsilon$ is $2 \pi$-antiperiodic function such that
\bee
\epsilon (p_{\mu}) =
  \left\{ \begin{array}{ll}
     1 \quad & \mbox{if  $p_{\mu} \in ( (4k-1) \pi,\: (4k+1) \pi)$}, \;
k \in \Z, \\
    -1 \quad & \mbox{otherwise}.
\end{array}
\right.
\ee
Although this constraint is nonlocal in the position space:
\bea
F^{\mu}_{n \: n'} &=&
\frac{1}{N^2} \sum_{p \in \Lambda^*} \exp [-i \:p\: (n-n'-
\frac{1}{2} \hat{\mu})] \: \epsilon (p_\mu) \cr
&=& \frac{1}{N} \: \frac{(-1)^{n_{\mu} - n'_{\mu} + 1}}
{\sin [\pi (n_{\mu} - n'_{\mu} - \frac{1}{2})/N]} \: \prod_{\nu \neq \mu}
\delta_{n_{\nu} \:
n'_{\nu}},
\eea
it does not bring a serious  problem, since the coupling of the fermions to
the gauge fields is local. Note that because of the remarkable property of
this constraint: $F = F^{-1}$, the propagator looks like a finite lattice
version of the SLAC formulation \cite{SLAC}.

Our second example is:
\bee
f^{\mu}(p) = \cos (\frac{1}{2} p_{\mu}).
\ee
In the case of (14) the constraint in position space looks most simple:
\bee
F^{\mu}_{n,\:n'} = \frac{1}{2} (\delta_{n,\:n'} - \delta_{n,\:n'+\hat{\mu}}),
\ee
and the propagator exactly coincides with the nonlocal formulation that is
dictated by the structure of the path integrals for the Weyl quantization
\cite{Ze}.

In both our examples the fermion propagators has no superfluous poles.

\section{Changes of variables}

The form of the path integral (8), being clear conceptually, is not very
convenient for practical calculations. Of course, one can get rid of the
constraint in the measure by introducing Lagrange multipliers, however it
does not simplify the problem.

Introduce new variables $\chi_n$ on the lattice sites such that
\bee
\chi_n = F^{0}_{n}[\chi^0],
\ee
and insert the identity $\int \prod_{n \in \Lambda} d \chi_{n} \:
\delta (\chi_n - F^{0}_{n}[\chi^0]) = 1$
into (8). Then, integrating over $\chi^0$ and $\chi^1$ we
come to the path integral with the measure
\bee
d \mu  =
\prod_{n \in \Lambda} d \chi_n  \: d \chi^{*}_{n}
\ee
and with the action
\bea
A_F[\chi^{*}, \chi; U]
= \sum_{n \in \Lambda} &\chi^{*}_{n}&
\left[ U_{n, n + \frac{1}{2}\hat{0}} \:
(F^{0})^{-1}_{n}[\chi]
-  U_{n, n - \frac{1}{2}\hat{0}} \:
(F^{0})^{-1}_{n-\hat{0}}[\chi] \right.
\cr
&&\mbox{} + i\: \left. (  U_{n, n + \frac{1}{2}\hat{1}} \:
(F^{1})^{-1}_{n}[\chi]
-  U_{n, n - \frac{1}{2}\hat{1}} \:
(F^{1})^{-1}_{n-\hat{1}}[\chi]
) \right].
\eea

By the definition (16) the new variables
have no simple transformation properties under the gauge group. However, if
we redefine the transformation such that $\chi_n \rightarrow h_n \:
\chi_n$, we come to
a perfectly conventional formulation with gauge invariant measure and
noninvariant action.
Now in the free field case $U = 1$ the action (18) (not only the propagator)
exactly coincides with the SLAC action \cite{SLAC} for $F$ from (13), and with
the Weyl action \cite{Ze} for $F$ from (15). The point by which the action
(18) differs from the preceding formulations, is the way by which the gauge
variables
enter it. It is important, that despite the fact that action in both cases
is nonlocal,
the coupling of the gauge fields to the fermions is local.

Note that there exists another change of the variables which is in a
certain sense remarkable, too. Define the variables $\chi'
_{n + \frac{1}{2} \hat{0} + \frac{1}{2} \hat{1}}$ on the centres
of the lattice plaquettes as follows:
\bee
\chi'_{n + \frac{1}{2} \hat{0} + \frac{1}{2} \hat{1}} =
(F^{1})^{-1}_n [\chi^0].
\ee
Then, following the same procedure as before, we get the measure
\bee
d \mu'  =
\prod_{n \in \Lambda} d \chi'_{n + \frac{1}{2} \hat{0} + \frac{1}{2} \hat{1}}
\: d \chi^{*}_{n}
\ee
and the action
\bea
{A'}_F[\chi^{*}, \chi'; U]
= \sum_{n \in \Lambda} &\chi^{*}_{n}&
\left[ U_{n, n + \frac{1}{2}\hat{0}} \:
F^{1}_{n}[\chi']
-  U_{n, n - \frac{1}{2}\hat{0}} \:
F^{1}_{n-\hat{0}}[\chi'] \right.
\cr
&&\mbox{} + i\: \left. (  U_{n, n + \frac{1}{2}\hat{1}} \:
F^{0}_{n}[\chi']
-  U_{n, n - \frac{1}{2}\hat{1}} \:
F^{0}_{n-\hat{0}}[\chi']
) \right].
\eea
Now both the measure and the action are not gauge invariant. However in case
(15), due to locality of the matrix $F$, action (21)
is local\footnote{Actions similar to (21) but with a different way of
introducing the gauge interactions were considered in ref. \cite{Tr}.}.

\section{Perturbative test}

In this section we examine in the perturbation theory the continuum limit of
the functional
$Z_F[U; 0, 0]$ with a smooth gauge field $U \in U(1)$ for variants (11) and
(15) of our formulation.

In the continuum theory the
perturbative solution to this problem is known to be exact \cite{Ano,JR}:
\bee
W[A] = -\ln  Z[A; 0, 0] =
\frac{1}{2} \int \frac{d^2 q}{(2 \pi)^2} A_{\mu}(-q) \:
\Pi_{\mu \nu}(q) \: A_{\nu}(q),
\ee
where $\Pi_{\mu \nu}$ is polarization operator of the field $A$:
\bee
\Pi_{\mu \nu}(q)
= \frac{e^2}{2 \pi} \left[ c \:
\delta_{\mu \nu} - \frac{q_{\mu}  q_{\nu}}{q^2}
+ \frac{i}{2 q^2} \: (\varepsilon_{\mu \alpha} q_{\alpha} q_{\nu} +
q_{\mu}  \varepsilon_{\nu \alpha} q_{\alpha}) \right].
\ee
Here $e$ is the gauge coupling,
$\varepsilon$ is the antisymmetric tensor, and $c$ is a parameter dependent
on regularization. The imaginary part of the effective
action $W$ is anomalous.

In terms of dimensionful variables the
limit we are interested in is $a \rightarrow 0$,
$q =$ const.. Calculate first the polarization operator.  Introducing
the gauge field
$A_{\mu}$, so that $U_{n, n \pm \frac{1}{2}\hat{\mu}} = \exp [(\pm i e a
A_{\mu}(n \pm \frac{1}{4}\hat{\mu})/2]$,  we have in this limit
\bea
\Pi_{\mu \nu}(q) = \lim_{N \rightarrow \infty}
\frac{1}{N^2} \sum_{p \in \Lambda^*} &[& V_{\mu}(p + q, -q) S(p + q)
V_{\nu}(p, q) S(p) \cr
& & \mbox{}- 2\: V_{\mu \nu}(p, q, -q) S(p) ],
\eea
where the propagator $S$ is defined in (10) and the vertices read as
follows:
\bea
V_{\mu}(p, q) &=& i\: e \: \sigma_{\mu} \cos (\frac{1}{2} p_{\mu} +
\frac{1}{4}
q_{\mu}) \: \frac{1}{f^{\mu}(p)}, \cr
V_{\mu \nu}(p, q, q') &=& - i\: e^2  \sigma_{\mu} \: \delta_{\mu \nu}
\frac{1}{4}
\sin (\frac{1}{2} p_{\mu} + \frac{1}{4} q_{\mu} + \frac{1}{4} q'_{\mu})
\: \frac{1}{f^{\mu}(p)},
\eea
with $\sigma_0 =1$, $\sigma_1 =i$. By direct computation of the sum in (24)
for increasing $N$ we find that for both variants of the $f^{\mu}$ the
expressions converge to the continuum form (23), but with
different values of constants $c$:
$c = 1$ in the case of (11) and $c \approx 1.28$ in the case of (14).

An analysis similar to that of ref. \cite{BK} shows that in the above limit
all the diagrams with the number of external legs not equal to 2 vanish.
Therefore, these results yield the exact answers for the fermion
determinats in a smooth external field.

The fact that for variant (11) one has $c = 1$ is remarkable.
Indeed, it shows that
in this case the gauge invariance is restored in the continuum limit for
the nonanomalous part of the fermion determinant without need of counterterms.

Meanwhile, the point that in  this formulation the gauge variables are defined
on the halfs of the lattice links leads to  some specific consequences.
Indeed, in this case the gauge field momenta $q_{\mu}$ belong to the interval
$(-2 \pi/a, 2 \pi/a]$
(in contrast
to the conventional case where $q_{\mu} \in (- \pi/a, \pi/a]$). Since the
momenta
on the lattice are conserved modulo $2 \pi/a$, in addition to the vertices
with the momenta $q_{\mu}$, there appear vertices with the
momenta $q_{\mu} + 2 \pi/a$. They, particularly, cause a violation of
Furry's theorem. In the continuum limit at $q$ are fixed, however, all
convergent diagrams with such vertices vanish, so that the only such diagram
that survives is that with single external momentum $q =
(2 \pi/a, \: 0)$ or $(0, \: 2 \pi/a)$. The consequences
of that for the full theory is to be studied, but it worth noting that such
vertices are cancelled if we introduce to the theory the Weyl fermions of
the same chirality but with variables $\psi$ and ${\psi^*}^{0}$,
${\psi^*}^{1}$ defined on the lattice sites and on the corresponding links,
respectively. At the same time the similar introduction of the fermions of the
opposite chirality, that may seem to be attractive taking in mind the gauge
invariant definition of the mass term: $\chi^* \psi + {\psi^*}^{0}
 \chi^0 + {\psi^*}^{1} \chi^1$, cancels such vertices only in the real
parts of the diagrams. In such a case the lattice Dirac operator is
complex, although, of course, the anomaly in the continuum limit is cancelled.

\section{Discussion}

We have demonstrated that in two dimensions variant (11) of our formulation
leads to the restoration of the gauge invariance of the nonanomaly
part of the fermion determinant.
The generalization to four dimensions can be done
straightforwardly, and we expect that the same features will hold in this
case, too. Although the fermion action is nonlocal, the gauge fields coupled
to the fermions in the local way, and therefore the formulation does not suffer
from the pathologies caused by nonlocal interactions.
The local variant of our formulation does not lead to the
restoration of the gauge invariance\footnote{Note, that a similar situation
arises as well in the Zaragoza proposal \cite{Za}.
In its standard form it is local and gauge noninvariant. However,
choosing {\it ad hoc} the form factor
suppressing the coupling of the gauge fields to the undesirable fermion modes
in a special step-like form (that is well nonlocal in the
position space), one can achieve restoration of gauge invariance in
the continuum limit (see the first reference in \cite{Za}). Such restoration
occurs also in the formulation with infinitely many fermionic degrees of
freedom \cite{NaN,AL}.}.
However, it is more economical than
formulations with the Wilson fermions, and can be implemented in the
regularization of the anomaly free chiral theories employing auxiliary
Pauli-Villars fields \cite{FS}.

The obvious drawback of this formulation is its gauge noninvariance at a
finite lattice spacing, even when it is applied to anomaly free models.
The perfect definition of the Weyl fermions would reproduce the anomaly but
keep nonanomalous part of the determinant to be gauge invariant. Then, if the
fermion content is adjusted in such a way that the theory is anomaly
free, its gauge invariance would be guarantied.

Surprisingly, such a result  can be achieved at the level of regularization of
chiral fermion loops. Indeed, we get this, if in the propagator (10) and
the vertices
(25) we put $f^{\mu}(p) = 1$. Then the real part of (24) turns out to be
transverse
even at a finite lattice spacing, while the total expression reproduces in the
continuum limit the correct answer (23) with $c = 1$. In the formulation with
$f^{\mu}(p) = \epsilon^{\mu}(p)$ the gauge invariance is violated by the
behaviour of the propagator and the vertices near the boundary of the
Brillouin zone, when in the fermion loop  one has
$p_{\mu} + q_{\mu} \not\in (-\pi, \: \pi)$ and the mechanism responsible for
$2 \pi$-periodicity of the propagator starts working. In the case of this
`perfect' regularization the propagator has no definite $2 \pi$-periodicity.
Therefore, the actual structure of the first loop in eq. (24) becomes
more complicated: in the half of the loop corresponding to the propagator
$S(p)$ always the left-handed fermions run, but in another half
the handedness of the fermions is changed depending on whether the momenta
are in the Brillouin zone or not. The same happens in the naive formulation
if the domain of integration over the fermion momenta is narrowed up to
$p_{\mu} \in (- \pi/2,\: \pi/2)$.

The main problem is that such `perfect' regularization exists only
as a prescription for regularization of the fermion loops and
does not allow a nonperturbative treatment of the theory. The first step
would be construction of the action which generates the propagator
$S(p) = -i / [ 2 \sin (\frac{1}{2} p_0) + i \: 2 \sin (\frac{1}{2}p_1)]$.
Since this $S(p)$ is not a function on the torus $S^1 \times S^1$, such an
action
cannot be constructed from the bilinear forms
$\chi^{*}_{p} \: S^{-1}(p) \: \chi_p$
on it. The interesting point however is that such a propagator is
a $2 \pi$-antiperiodic function on the real projective plane $\R P^2$, which
is obtained from the square momentum lattice (see the footnote 4)
by the addition of links connecting the site $(\pi(N-1),\: p_1)$ with the
site $(-\pi(N-1),\: -p_1)$ and the site $(p_0,\:\pi(N-1))$ with
$(-p_0, \: -\pi(N-1))$.
Then the action can have the form $\sum_{p \in \Lambda^*} \chi^{*}_{p} \:
S^{-1}(p) \: \chi_p$, where now $\Lambda^*$ topologically is $\R P^2$ and
$\chi^*$ ( $\chi$ ), say, $2 \pi$-antiperiodic ($2 \pi$-periodic)
function on $\Lambda^*$.

\section{Acknowledgement}

I am grateful to E. Seiler, O. Ogievetsky, and  P. Breitenlohner for
enlightening discussions on the groups on $\R P^2$, to E. Seiler for
reading the manuscript and useful comments, and to H. Neuberger for
correspondence. It is a pleasure to thank
E. Seiler and the Theory Group of Max-Plank-Institut f\"ur Physik, where
the final part of this work has been done, for their kind hospitality.
This work was partly supported by the Russian Basic Research Fund under
grant 95-02-03868a.

\end{document}